\documentclass[a4paper]{article}

\usepackage{graphicx}
\usepackage{times}
\usepackage[hyphens]{url}
\usepackage[breaklinks]{hyperref}
\usepackage{breakurl}
\usepackage[a4paper]{geometry}

\begin{document}
\begin{titlepage}
\title{Investigation the connection between the intermediate gamma-ray bursts and X-ray flashes}
\date{}
\maketitle
\begin{center}\large
J\'ozsef K\'obori$^{1}$,
Zsolt Bagoly$^{1}$,
Lajos G. Bal\'azs$^{2,3}$ and
Istv\'an Horv\'ath$^{4}$ \end{center}
$^{1}$Dept. of Physics of Complex Systems, E\"otv\"os University, Budapest, Hungary, jkobori$@$caesar.elte.hu\\
$^{2}$MTA CSFK Konkoly Observatory, Budapest, Hungary,\\
$^{3}$Dept. of Astronomy, E\"otv\"os University, Budapest, Hungary,\\
$^{4}$Dept. of Physics, National University of Public Service, Budapest, Hungaryű
\normalsize\abstract{Gamma-ray bursts (GRBs) can be divided into three groups: short-, intermediate- and long-duration bursts.
While the progenitors of the short and long ones are relatively known, the progenitor objects of the intermediate-duration bursts (IBs)
are generally unknown, however, recent statistical studies suggest, that they can be related to
the long-duration bursts. Another types of GRBs are the so-called X-ray flashes (XRFs) and X-ray rich GRBs (XRRs).
The former ones radiate
more intensively in the X-ray bands than common GRBs, but in the cases of XRFs the main component of the emission is produced 
entirely at X-ray wavelengths. Also, the XRFs and IBs show some similarities regarding their prompt- and afterglow
properties.
In this work we investigate whether there is a difference between the global parameters of 
the X-ray flashes and intermediate-duration group of gamma-ray bursts.
The statistical tests do not show any significant discrepancy regarding most of the parameters, except the BAT photon index,
which is only a consequence of the defintion of the XRFs.}

\maketitle
\end{titlepage}

\section{Introduction}
Gamma-ray bursts are a mysterious phenomenon since their discovery in 1967. In 1993, Kouveliotou et al. (1993),
based on statistical studies showed that two types of GRBs can be distinguished using the $T_{90}$ quantity 
($T_{90}$ is the time during which the cumulative counts increase from 5\% to 95\% above the background, thus encompassing 90\% of the total GRB counts):
the short- ($T_{90} < 2$ s) and long-duration $T_{90} > 2$ s) bursts. Later on, Horv\'ath (1998) and Mukherjee et al. (1998) in 1998 independently have
found, that an intermediate-duration type of bursts exists as well.
These bursts can be connected to the short- (e.g. \v{R}ipa et al. 2012) and long-duration (e.g. Veres et al. 2010) 
bursts as well depending on the particular gamma-ray satellite used for constructing the sample.
Since then some other types of bursts were identified: X-ray rich bursts and X-ray flashes.  
Statistical analyses suggest (e.g. Sakamoto et al. 2008) that common-GRBs, XRRs and XRFs can be drawn from the same population.
However, the debate about the underlying physics is still going on, but thanks to the \textit{Swift} mission (Gehrels et al. 2004),
studying the simultaneous observations at many wavelengths allow us to constrain some
of the prompt- and afterglow-emission parameters for the various types of bursts.

In this paper we report a statistical study of the prompt and afterglow emission of the intermediate group
of GRBs and X-ray flashes. In \S 2 we describe the classification method of GRBs, then,
in \S 3 we investigate the global parameters of the BAT lightcurves and spectra.
After that, in \S 4 the XRT properties are analyised, while UVOT afterglows are presented in \S 5.
\section{The data sample and the classification method}

The sample used in this paper was created by Veres et al. (2010), who
classified the bursts using the model based clustering (Bayesian Information Criterion) method.
This method gives a probability that a burst belongs to a group, instead of a definite membership.
For details of the process, please see their article.
The sample consists of 408 bursts up to GRB090726: 331 long, 46 intermediate and 31 short bursts.
All of the bursts were detected by the Swift satellite.
The observational data was downloaded from the Swift Science Data Center
(BAT fluence, XRT 11- and 24-hour flux, XRT Column Density (NH), BAT 1-sec peak photon flux, BAT photon index, XRT spectral index)
and The Swift-XRT GRB Catalogue (temporal decay indices, break times).

Among the 46 intermediate bursts there are 38 bursts for which temporal decay indices are available in the XRT Catalogue,
but further 5 bursts are excluded, because their lightcurve consists of more than 2 breaks and one or more flares
(in these cases because of the flare activity it is difficult to fit the lightcurve, hence determine the underlying 
afterglow decay index). So, in the following calculations we analysed 33 bursts, 
and on the figures below 33 bursts are plotted as well.

If we apply the definition for XRFs adopted by Sakamoto et al. (2008) ($H_{32} < 0.76$,
where $H_{ij}$ ($i,j$ mark two energy intervals) is the ratio of the fluences in different channels for a given burst using the usual \textit{Swift} energy bands with
$10-25-50-100-150$ keV as their boundaries, so $H_{32} = \frac{S_{50-100}}{S_{25-50}}$), then we can find
20 XRFs and 13 IBs in our sample.

To test whether the XRFs and IBs are drawn from the same population we compared their emission properties applying
the Kolmogorov-Smirnov test to the data.

\section{BAT fluences, peak fluxes and photon indices}
\begin{figure}[!h]
\begin{center}
\includegraphics[angle=0, width=11cm]{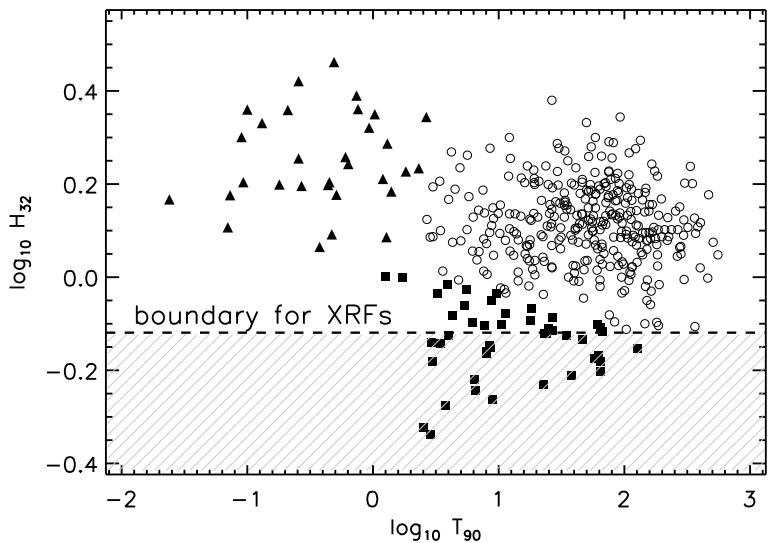}
\end{center}
\caption{On the figure we plotted the entire sample consisting the long (open circles), intermediate (filled squares) and short GRBs (filled triangles).
The dashed line marks the boundary for XRFs defined by Sakamoto et al. (2010), that is, if $H_{32} < 0.76$ the burst is indetified as an X-ray flash.}
\label{vp}
\end{figure}

\begin{figure}[!h]
\begin{center}
\includegraphics[angle=0, width=11cm]{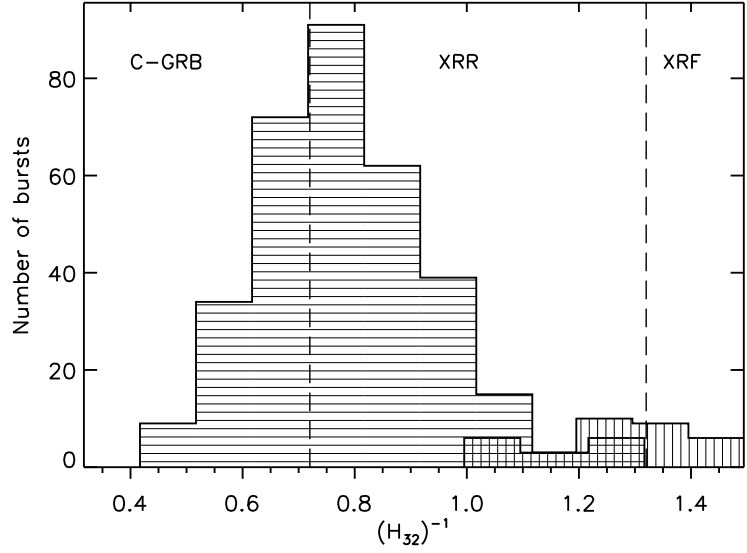}
\end{center}
\caption{Distributions of the fluence ratio S(25-50 keV)/S(50-100 keV) in our sample for long bursts (horizontal line area)
and intermediate bursts (vertical line area). The dashed lines correspond to the borders
between common-GRBs and XRRs, and between XRRs and XRFs.}
\label{sakamoto}
\end{figure}

A thorough analysis of the Swift BAT ($\gamma$-detector) and XRT (X-ray detector) data was carried out by Sakamoto et al. (2008), who 
investigated the spectral and temporal characteristics of the prompt- and X-ray afterglow-emission 
of XRFs and XRRs detected by \textit{Swift} between 2004 December and 2006 September.
They have found that XRFs, XRRs and the common GRBs 
form a single broad distribution.
For comparison, we constructed (Fig. \ref{sakamoto}) the histogram of the quantity $(H_{32})^{-1}$ of our sample for the long-duration bursts, 
and it confirms the previous result (common distribution for ordinary GRBs, XRFs and XRRs). 
Nevertheless, if we plot the $H_{32}$ against the $T_{90}$ for the entire sample (short, intermediate and long GRBs) (\ref{vp}) we can see that the boundary for XRFs overlap the range of IBs.

\begin{figure}[!h]
\begin{center}
\includegraphics[angle=0, width=11cm]{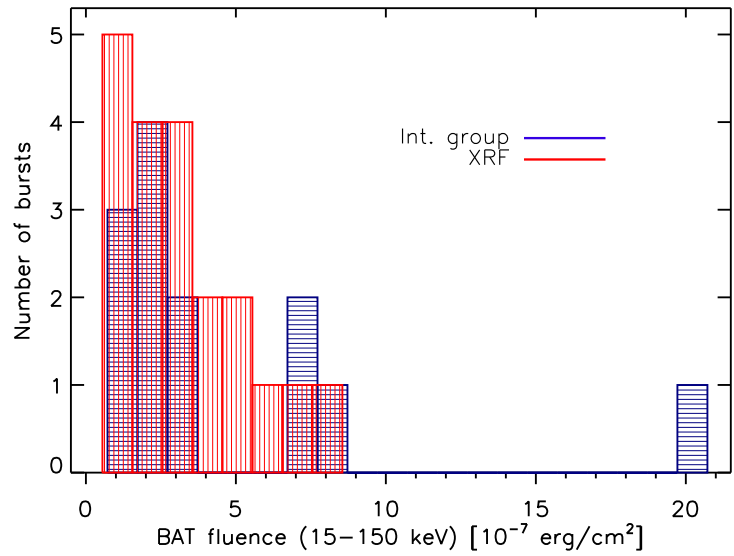}
\end{center}
\caption{The histogram shows the distribution of the BAT fluences for
XRFs and IBs. As it can be clearly observed, the XRFs and IBs form a uniqe population
regarding the BAT fluence.}
\label{batflu}
\end{figure}

\begin{figure}[!h]
\begin{center}
\includegraphics[angle=0, width=11cm]{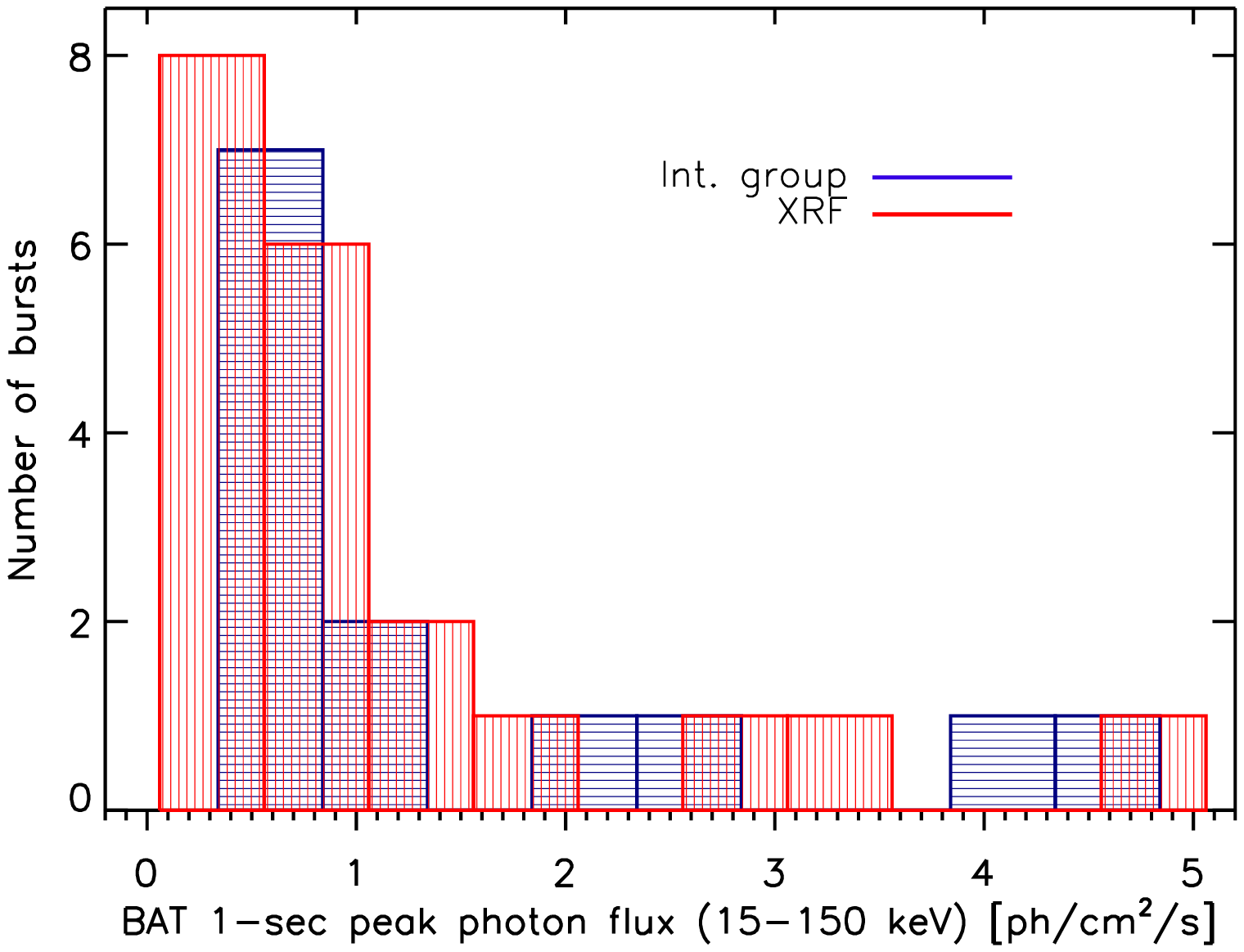}
\end{center}
\caption{On the figure the distribution of the BAT 1-sec peak photon fluxes can be seen.
The KS test indicates that the different types of bursts are drawn from
the same population.}
\label{batpeak}
\end{figure}

This result is discussed in details by Veres et al. (2010), and it suggests a connection between 
XRFs and IBs.
If we take a look at the histogram of BAT fluences measured between 15-150 keV (Fig. \ref{batflu}), we can see
that there is no difference in the distribution of XRF's and IB's BAT fluences. The KS test
supports this hypothesis with significance of 0.87. 
In the cases of the BAT 1-sec peak photon fluxes (Fig. \ref{batpeak}), the result is similar,
the null hypothesis is accepted with significance of 0.28.

Contrary to the similarities in the fluence and peak flux distributions, 
a remarkable discrepancy can be observed in the distribution of the BAT photon indices.
The reason of this bias is that the constrain for the XRFs ($H_{32} < 0.76$) was defined arbitrarily,
which means, that if we choose a higher value for $H_{32}$ as the boundary, then the difference in the distribution
simply disappears.
\begin{figure}[!h]
\begin{center}
\includegraphics[angle=0, width=11cm]{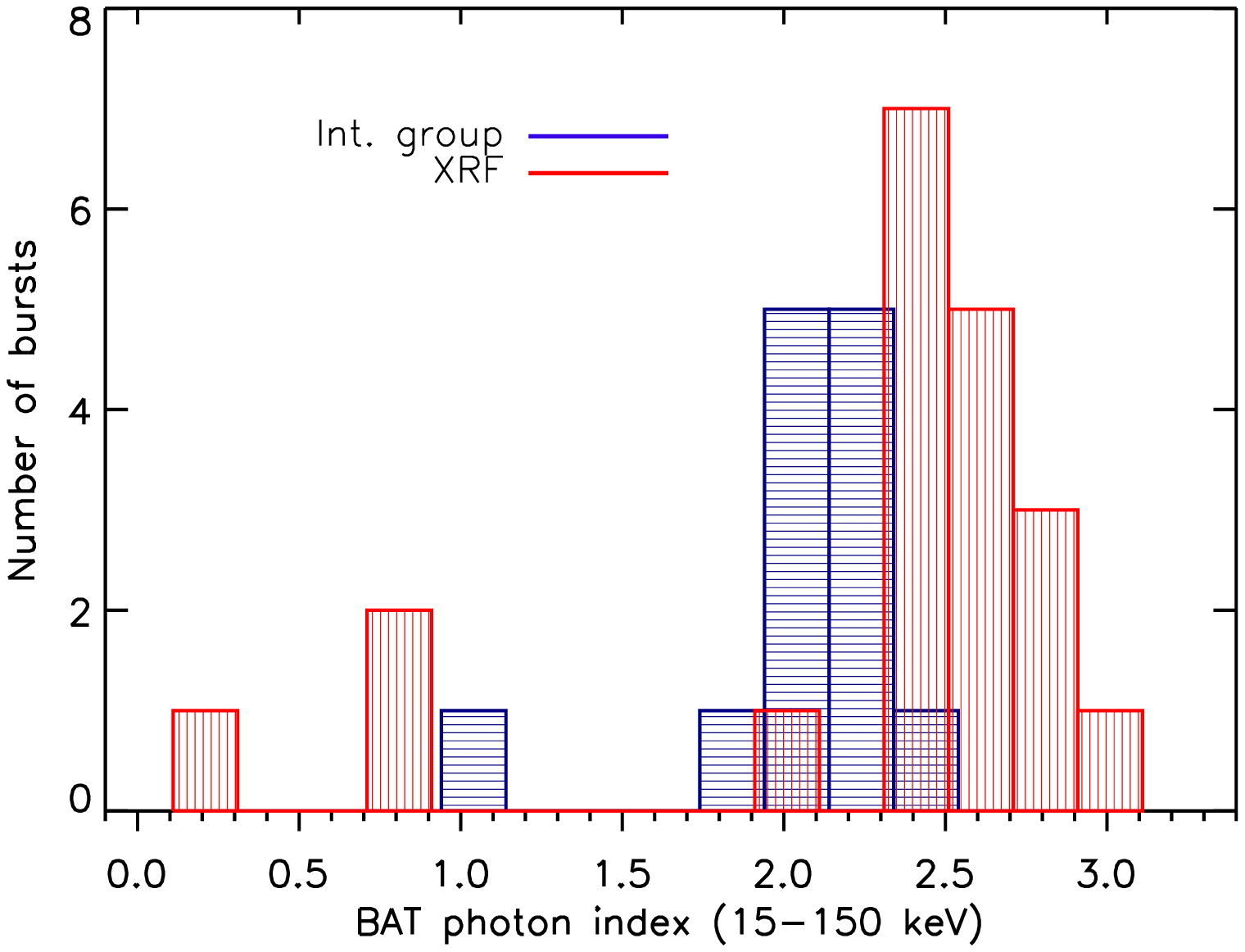}
\end{center}
\caption{If we examine the distribution of the BAT photon index, a marginal distinction
can be revealed between the XRFs and IBs,
but it is not suprising, since the boundary for the XRFs ($H_{32} < 0.76$) was defined
arbitrarily. If we set this value higher, the significance of the discrepancy becomes lower and finally disappears.}
\label{phi}
\end{figure}
\section{XRT temporal and spectral indices}
\begin{figure}[!h]
\begin{center}
\includegraphics[angle=0, width=11cm]{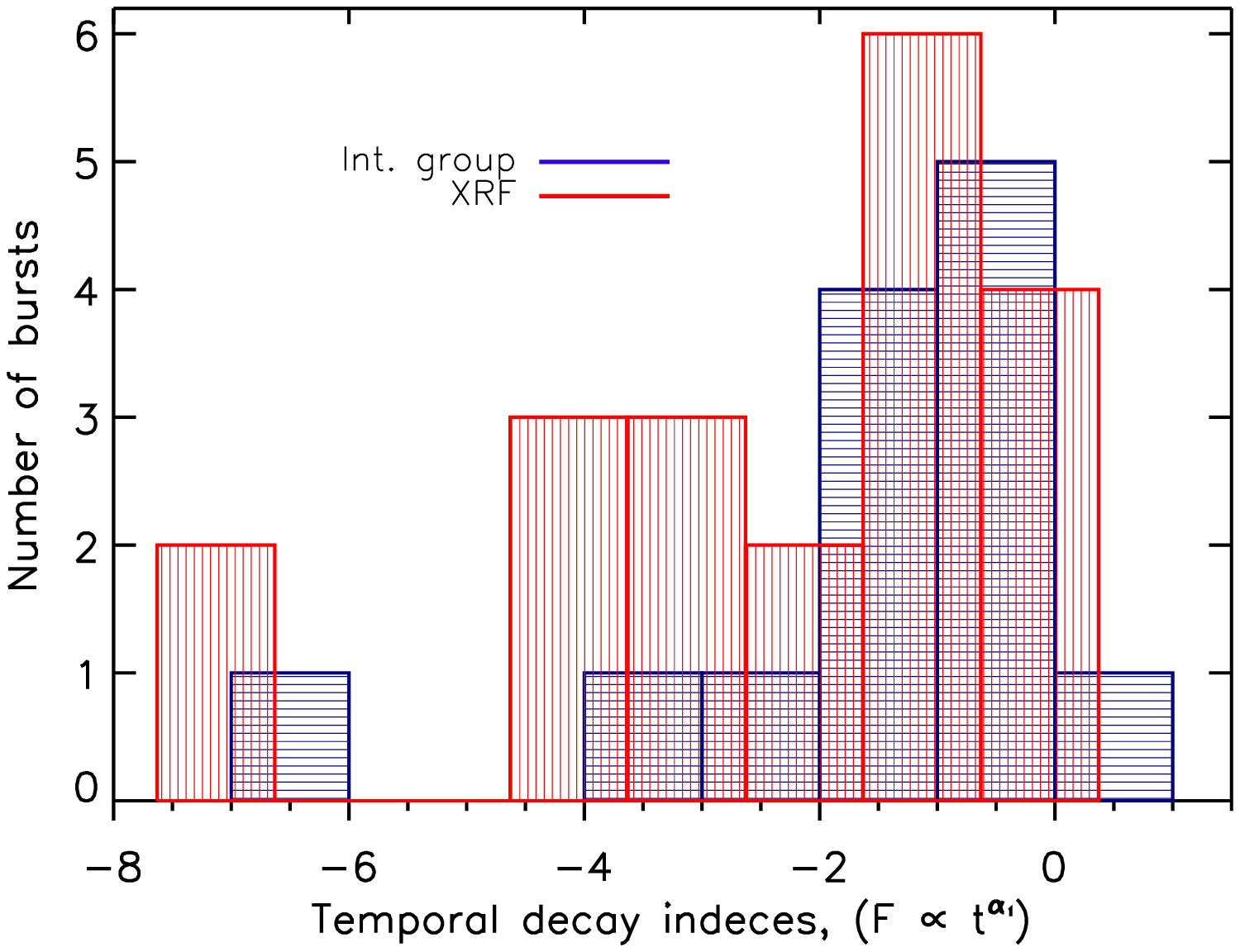}
\end{center}
\caption{The initial ($\alpha_1$) XRT decay indices are similar for the XRFs and IBs. The 
KS test does not suggest any difference between the distributions (the significance for having the 
same parent distribution hypothesis is 0.35).}
\label{x1}
\end{figure}

\begin{figure}[!h]
\begin{center}
\includegraphics[angle=0, width=11cm]{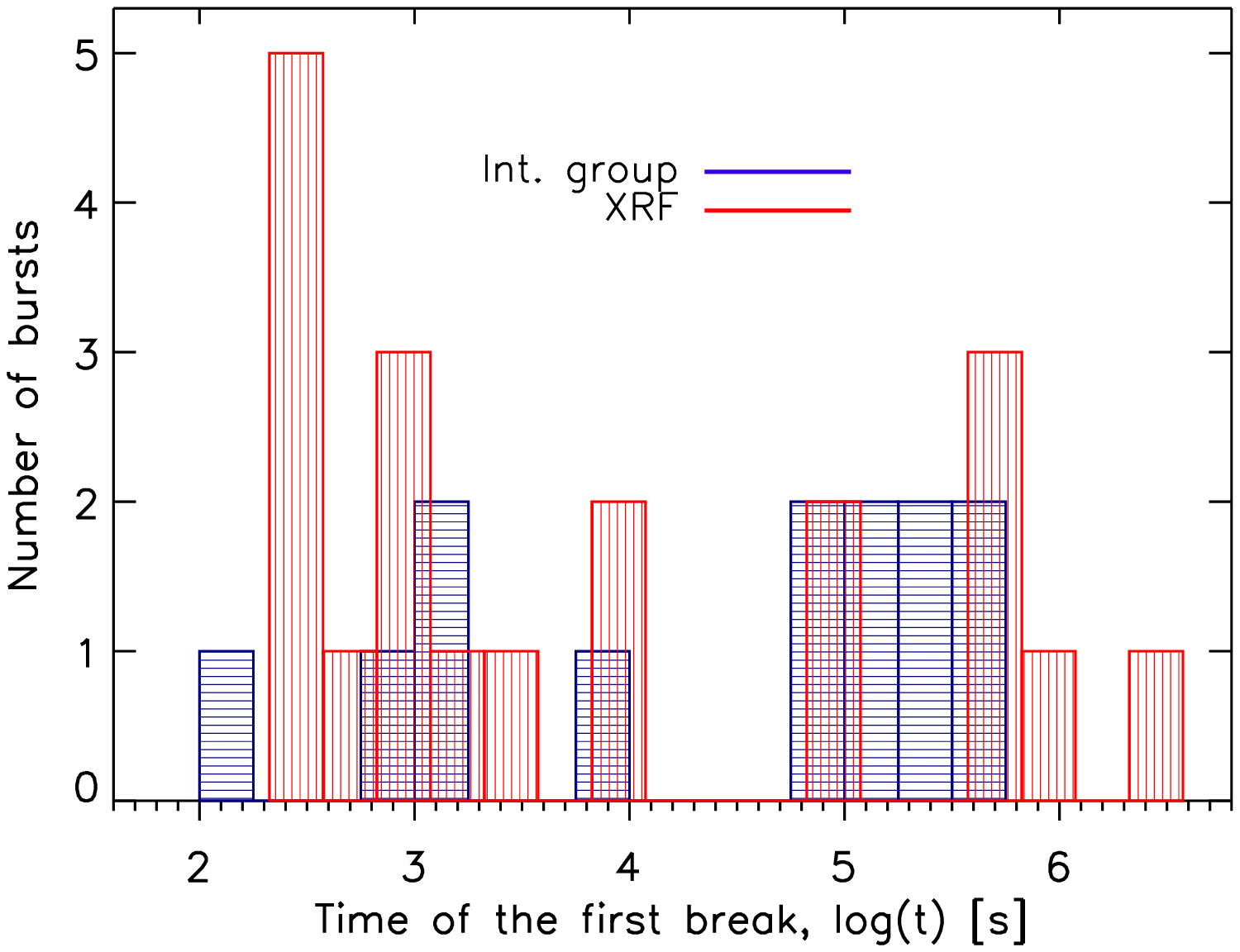}
\end{center}
\caption{The figure clearly shows, that the first break in the XRT lightcurves
appears to happen earlier for the majority of the XRFs, although,
the KS test gives a probability only of 89\% for having different populations.}
\label{tbreak1}
\end{figure}

We also compared the distributions of XRT afterglow properties of the XRFs and IBs.\\
On Fig. (\ref{x1}) we plotted the initial temporal indices, which characterise the first power-law
segment of the XRT afterglows.
In the distribution of the first break time (Fig. \ref{tbreak1})
a slight difference can be
discovered, the X-ray flashes tend to have this break earlier than the intermediate bursts.

\begin{figure}[!h]
\begin{center}
\includegraphics[angle=0, width=11cm]{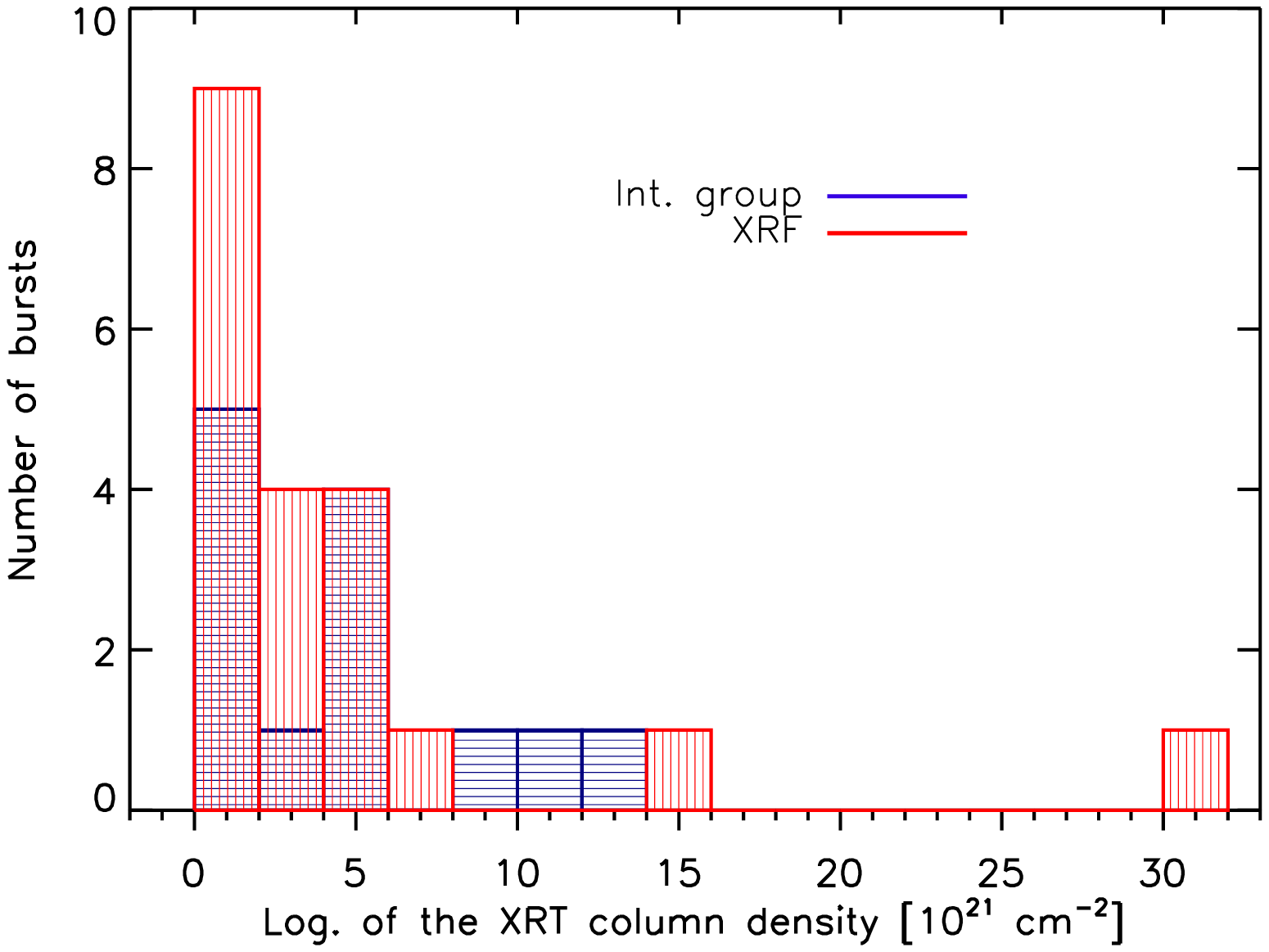}
\end{center}
\caption{As the  majority of other prompt and afterglow parameters, the 
neutral hydrogen column densities of XRFs and IBs form a single population.}
\label{nh}
\end{figure}

Investigating the $\alpha_1$ and $\alpha_2$ indices ($F \propto t^{-\alpha_i}$), the similarity between the XRFs and IBs still holds on, 
the significances for being drawn from the same populations are 0.42 and 0.35, respectively.
\begin{figure}[!h]
\begin{center}
\includegraphics[angle=0, width=11cm]{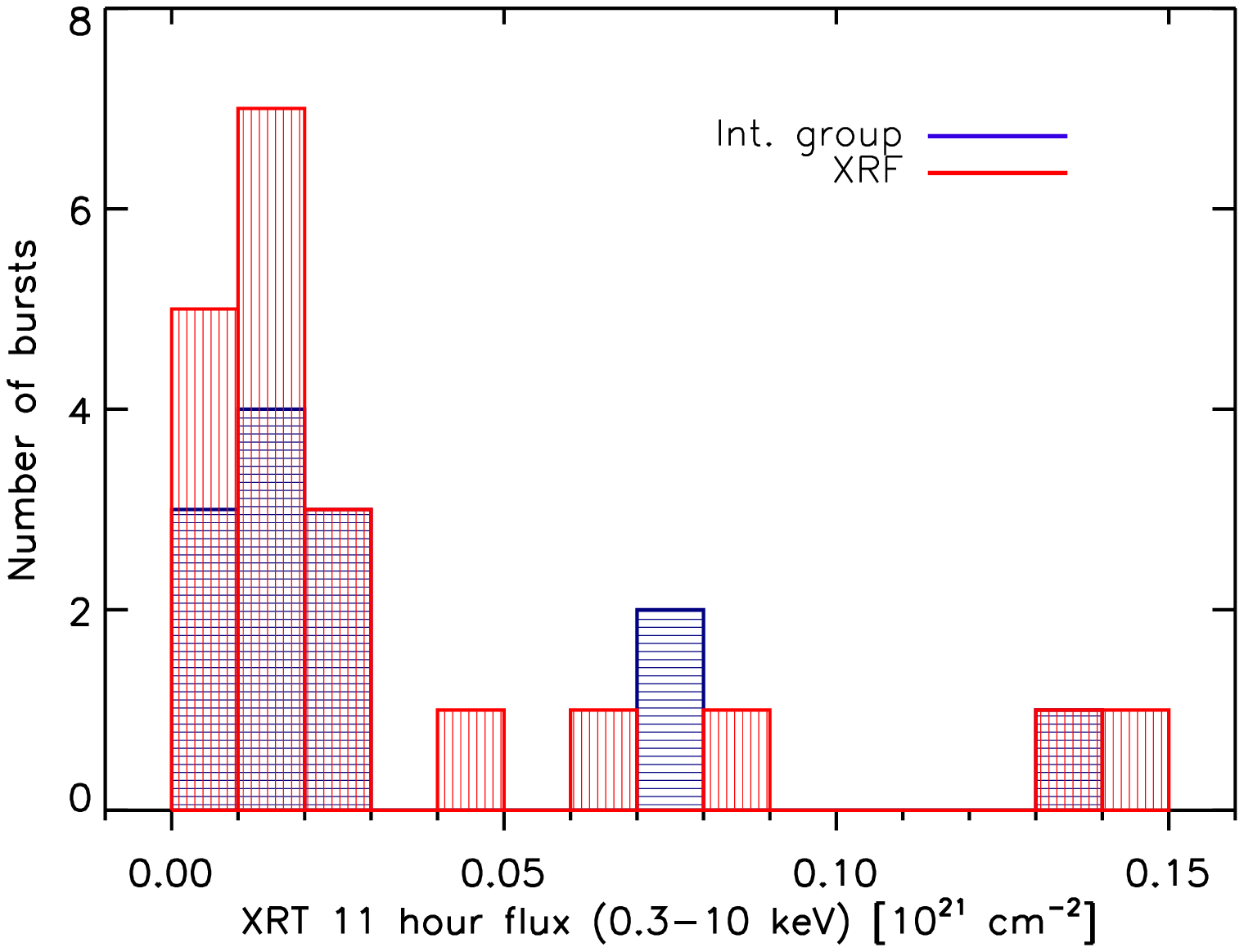}
\end{center}
\caption{The distribution of the XRT fluxes measured at 11 hours after the trigger confirms
the previous results.}
\label{x11}
\end{figure}
The KS test also accepts the same population hypothesis for the XRT fluxes
measured at 11 and 24 hours after the trigger with probabilities of 28\% and 11\%, which is not surprising,
since the $\alpha_1, \alpha_2$ indices form the same distribution as well.
The distributions of the neutral hydrogen column densities (Fig. \ref{nh}) in the host galaxies calculated from the XRT afterglows
indicate that the XRFs and IBs tend to occur in similar type of galaxies, this
is confirmed by the KS test, the same population hypothesis has significance of 0.43.

\begin{figure}[!h]
\begin{center}
\includegraphics[angle=0, width=11cm]{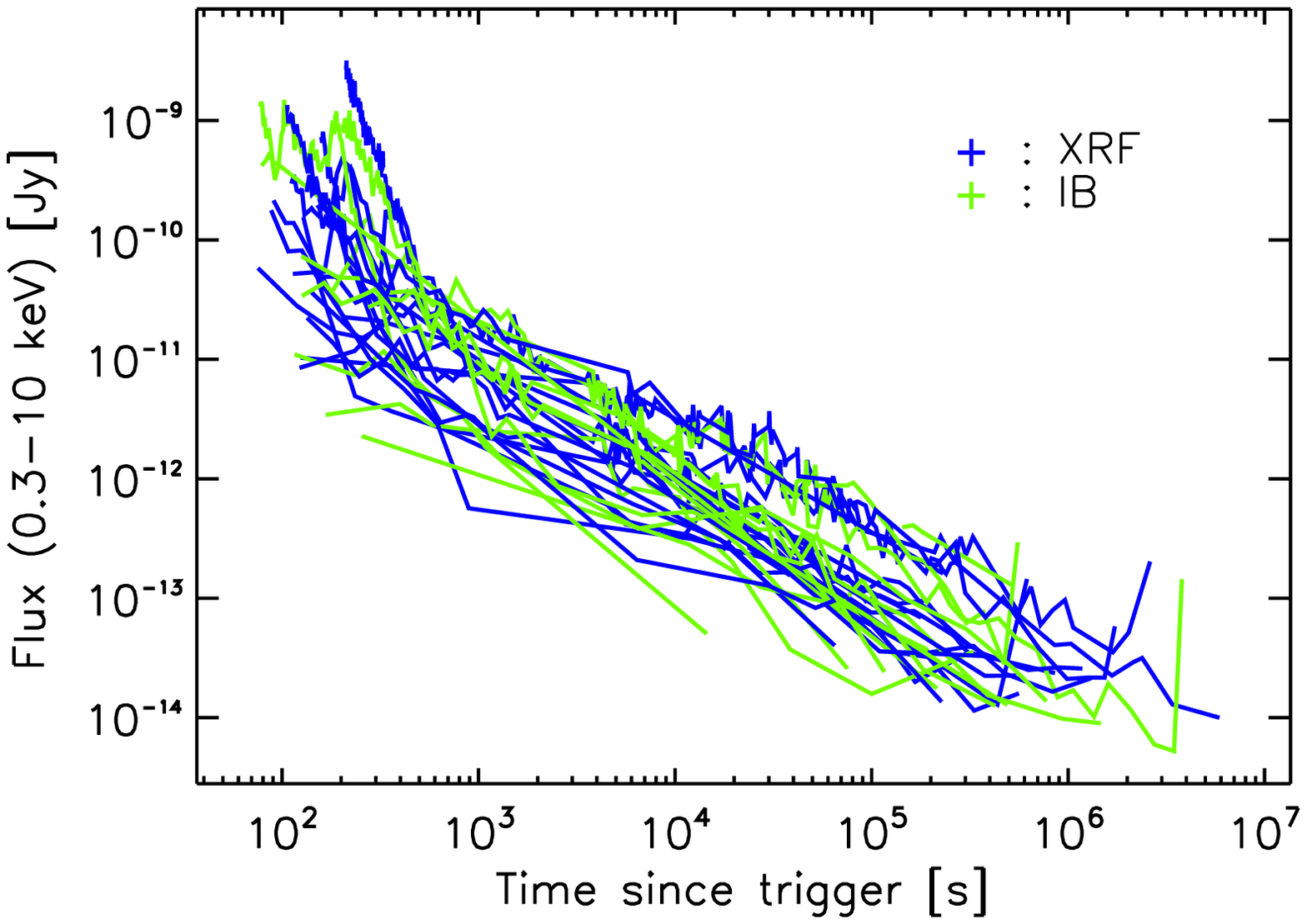}
\end{center}
\caption{As it can be observed, most of the XRFs have a lightcurve with steeper initial temporal phase,
but after the first break, there is no difference in the decay index distribution.}
\label{xrtlc}
\end{figure}

Finally, we constructed the XRT lightcurves of the XRFs and IBs (Fig. \ref{xrtlc}). As the distribution
of the initial temporal decay indices suggests (Fig. \ref{x1}), the XRF afterglows have a steeper
initial phase, but after the first break the decay rates are consistent between the two groups.
\section{UVOT afterglows}
From the 46 IB, 15 has UVOT observations, but only 11 of them is bright enough
to construct the optical lightcurve. After applying the criterion for XRFs, it turns out, that
from the 11 afterglows 4 belong to the IBs and 7 belong to the XRFs.\\
\begin{figure}[!h]
\begin{center}
\includegraphics[angle=0, width=11cm]{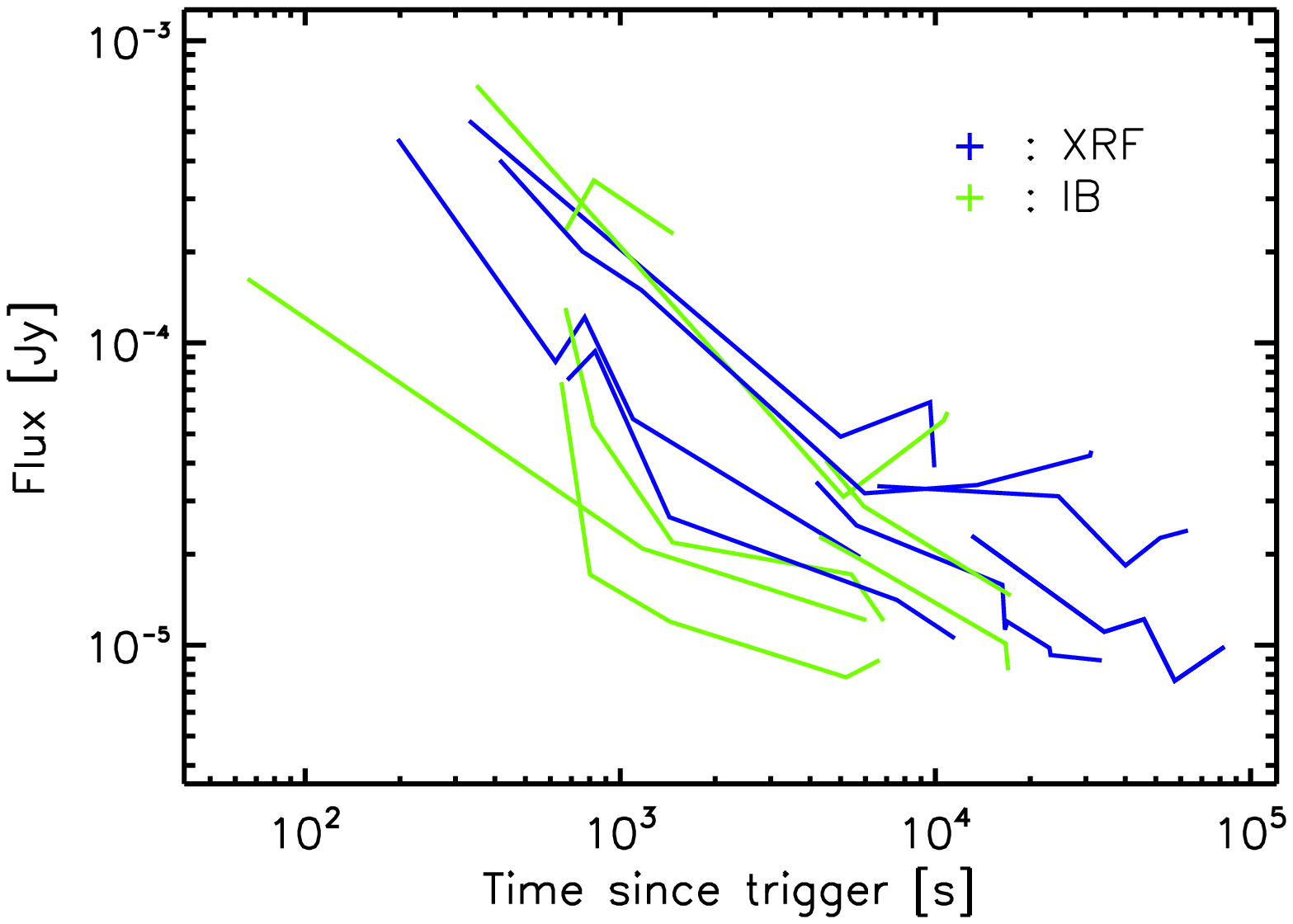}
\end{center}
\caption{On the figure the UVOT (V, B and U) lightcurves of XRFs and IBs can be observed.
The figure does not suggest any discrepancy among the XRFs and IBs.}
\label{lc}
\end{figure}
Regarding the optical lightcurves there is not any difference between
the lightcurves of XRFs and IBs, they lie in the same flux range (Fig. \ref{lc}).
\section{Conclusion}
Veres et al. (2010), based on the $H_{32}$ and $T_{90}$ distributions, showed that the IBs and XRFs probably have the same physical origin,
the observable differencies in the $\gamma-$, X-ray and optical bands may be the consequencies of the minor changes in the underlying
physical processes, circumburst medium properties and/or observational effects. 
Our main result in this article that beside the above mentioned $T_{90}$ and $H_{32}$ the IBs and XRFs show
similar properties regarding the other observed ($\gamma-$, X-ray and optical) quantities detected by the \textit{Swift} gamma-ray satellite.
We have to emphasize that our conclusion concerns GRBs observed \textit{only} by the \textit{Swift}.
\section*{Acknowledgements}
This research has made use of data obtained through the High Energy Astrophysics 
Science Archive Research Center Online Service, provided by the NASA/Goddard Space Flight Center
and the data supplied by the UK Swift Science Data Centre at the University of Leicester.
This study was supported by the Hungarian OTKA grant No. T77795.

\end{document}